# Mesoscopic model for colloidal particles, powders, and granular solids

Robert D. Groot and Simeon D. Stoyanov
*Unilever Research Vlaardingen,*
*P.O. Box 114, 3130 AC Vlaardingen, The Netherlands*

A simulation model is presented, comprising elastic spheres with a short range attraction. Besides conservative forces, radial- and shear friction, and radial noise are added. The model can be used to simulate colloids, granular solids and powders, and the parameters may be related to experimental systems via the range of attraction and the adhesion energy. The model shares the simplicity and speed of Dissipative Particle Dynamics (DPD), yet the predictions are rather non-trivial. We demonstrate that the model predicts the correct scaling relations for fracture of granular solids, and we present a schematic phase diagram. This shows liquid-vapor coexistence for sufficiently large interaction range, with a surface tension that follows Ising criticality. For smaller interaction range only solid-vapor coexistence is found, but for very small attractive interaction range stable liquid-vapor coexistence reappears due to pathological stability of the solid phase. At very low temperature the model forms a glassy state.

PACS number(s): 82.70.-y, 64.75.Gh, 45.70.-n, 62.25.Mn

## I. INTRODUCTION

In the last ten years attention has been drawn toward the use of mesoscopic simulation methods for liquid systems like polymer melts, bio-membranes and surfactants, where the system of interest has a relevant length scale between the atomistic scale and the macroscopic scale. A popular simulation method for such mesoscale problems is Dissipative Particle Dynamics (DPD) [1,2]. One shortcoming of DPD is that it is suited only to simulate liquids; another shortcoming is its failure to reproduce liquid-vapor equilibrium. The aim of the present paper is to fill this gap and formulate the simplest extension that is capable to simulate liquids and solids, while retaining much of the simplicity and simulation speed of DPD.

Although many problems in soft condensed matter can be addressed by a fluid particle model, in some areas the solid character of the particles is essential. One example is the stabilization of emulsions or foams by colloidal particles [3]. Foam stabilization by solid particles is beneficial for a number on reasons. A first reason is the high surface adsorption energy of particles, but particle shape [4] and particle-particle interactions also play crucial roles as pointed out by Binks [5]. Hence, for this problem it is essential to gain understanding of the adsorption and phase behavior of (mixed) colloidal systems, in dependence of the range and scale of the inter-particle interactions.

Another example where the solid character of particles is important is in the flow and stability of granular powders and solids. A close link between powders and liquids was established by Edwards [6], who showed that standard statistical mechanics can be used to describe powders, by drawing an analogy between temperature and mechanical vibrations. This way, the jamming transition in powder flow has been related to the glass transition in liquids [7]. In some simulation studies purely repulsive particles are used [8,9], others use very steep Lennard-Jones like potentials [10]. In general, capillary forces lead to a short-range attractive interaction in wet powders [11] which determine the phase behavior. In particular, very different regimes of granular media have been distinguished that correspond to glassy, liquid and granular gas phases [12]. Thus, for powders too it is important to know the phase behavior as function of the interaction range and strength.

For colloidal particles Hagen and Frenkel [13] found that if the interaction range is shorter than about 1/6th of the particle diameter, liquid-vapor coexistence disappears. On the other hand, Miller and Frenkel [14] successfully simulated liquid-vapor coexistence for the Adhesive Hard Sphere model of vanishing interaction range. These results seem to be in conflict. To better understand powder rheology, colloid stability, and foam stabilization by particles, we thus need to revisit the phase behavior as function of temperature and interaction range for adhesive particles. To this end, the simplest possible simulation model has been studied, which will be outlined in the next section. The phase diagram is presented in section III, and conclusions are formulated in the last section.

## II. INTERACTION MODEL AND SIMULATION METHOD

### A. Conservative forces

The standard DPD model is based on a linear elastic spring interaction potential, typically with a maximum repulsion force $a$ = 25 – 100 $kT/d$, where $d$ is the particle diameter [2]. A linear elastic spring is clearly the simplest model to represent liquid elements. This model is however unsuited to simulate solids; the main reason is the softness of the interaction potential. Indeed, Stoyanov and Groot [15] mention the formation of soft solids for $ad > 500$ $kT$. Hence, the simplest model to simulate solids is to use linear elastic springs with a high repulsion parameter. However, this model cannot reproduce liquid-vapor or solid-vapor coexistence, for which a cohesive force is needed. The simplest possible extension to the DPD model that will allow liquid-vapor and solid-vapor coexistence is therefore to use elastic spheres with a short-range attraction.

To obtain a reasonable form for the repulsive and attractive forces, we first study two elastic spheres of diameter $d$ (or radius $R$). In the simplest approximation we assume affine deformation and volume conservation. The elastic energy per particle is then given by $W_e = \frac{1}{6}VE(\Sigma_i \lambda_i^2 - 3)$, where $V$ is the particle volume, $E$ is the linear elastic modulus, and the $\lambda_i$s are the principal deformation ratios. If the spheres are deformed in the $z$-direction by a factor $\lambda_z = 1-u/d$, the deformations in $x$ and $y$-directions are given by $\lambda_x = \lambda_y = 1/(1-u/d)^{1/2}$. Hence the elastic energy is $W_e = 2\pi R^3 E [(1-u/d)^2 + 2/(1-u/d) - 3]/9$. The force needed for this deformation is found as F = $-\partial W_e/\partial u \approx 2\pi R^2 Eu/3d$, or

$$F^{Rep}(r) = a\left(1 - \frac{r}{d}\right) \quad (r < d) \quad (1)$$

where the pre-factor $a$ is given by $a = 2\pi ER^2/3$, and $r$ is the distance between particle centers. Note that this is the usual choice for the repulsive force in DPD, where the front factor is related to the modulus of the particle.

At separation distances further away than a particle diameter, the force should become attractive because particles at contact form hydrogen bonds, have a hydrophobic interaction, or have a solid or liquid bridge. The simplest form to represent this, is to assume a parabola force,

$$F^{Att}(r) = \frac{4\varepsilon}{\delta^2}(1 - r/d)(1 + \delta - r/d) \quad (2)$$

Here, $\delta$ is the range of the attractive interaction relative to the particle diameter and $\varepsilon$ is the force minimum. We will refer to the present model as the Sticky Elastic Sphere model.

To simulate colloidal particles that are e.g. bound together by hydrogen bonds, one might be tempted to take an extremely small value for $\delta$, because the range of a hydrogen bond is very small. However when a bond is formed, the distance between particle centers may still vary even if the surfaces are glued together with no intervening space, if the particles are elastic. Therefore, the range of the attractive interaction is actually set by the elastic modulus of the particles, and not by the range of the hydrogen bond.

To find the connection to a particular physical system we first remark that at $r = d$ the derivative of the force must be continuous, which implies the following relation between the force amplitude, the force range and the modulus

$$\varepsilon = \tfrac{1}{4}a\delta \quad (3)$$

The interaction forces in Eqs. (1) and (2) correspond to the interaction potential

$$U(r) = \begin{cases} U^{Rep}(r) - \frac{2}{3}\varepsilon\delta d & \text{for } r < d \\ \frac{4\varepsilon d}{\delta^2}\left[\frac{1}{3}(1+\delta-r/d)^3 - \frac{1}{2}\delta(1+\delta-r/d)^2\right] & \text{for } d < r < (1+\delta)d \end{cases} \quad (4)$$

where $U^{Rep} = \frac{1}{2}ad(1-r/d)^2$ is the interaction energy of the purely repulsive interaction force, Eq. (1). The energy minimum, at $r = d$, follows as

$$-U^{min} = G = \frac{2}{3}\varepsilon\delta d \quad (5)$$

Combining Eqs. (3) and (5), we can eliminate the force amplitude, and obtain the interaction range as

$$\delta = \sqrt{\frac{6G}{ad}} = \sqrt{\frac{6G}{EV}} \quad (6)$$

where $V$ is the particle volume. It is easily checked that the right hand side expression in Eq. (6) is dimensionless since $G$ is an energy, and $E$ is a modulus. Hence, the range $\delta$ is a proper dimensionless group that should match between simulation and experiment.

Next we need to choose our unit of energy. The relevant dimensionless group is the adhesion energy $G$ divided by thermal energy $kT$, $G^* = G/kT$, as this determines the association constant between particles. Once the adhesion energy, modulus and particle volume are known, the interaction range is fixed and can be used as simulation parameter. Using Eq. (6) we can express the slope of the repulsive force in Eq. (1) in terms of $G$ and $\delta$, and find

$$a = \frac{2\pi}{3}ER^2 = \frac{3G}{\delta^2 R} = EV/d \quad (7)$$

In the simulation we choose the particle diameter as a length scale. For Brownian particles we can define the energy scale such that $kT = 1$, the dimensionless repulsion parameter then follows as

$$a^* = \frac{ad}{kT} = \frac{EV}{kT} \quad (8)$$

Substitution of Eq. (7) into Eq. (3) now gives the force amplitude as

$$\varepsilon = \frac{3G}{4\delta R} = \frac{3}{4R}\sqrt{EVG/6} \quad (9)$$

Even though $\delta$ is a practical parameter from a simulation point of view, in reality the modulus and particle size will be fixed when the association energy is changed by altering salt concentration, pH or temperature. In that case parameter $a$ should be kept constant. For typical solid polymer particles of 20 nm diameter and modulus 2 GPa, this would give a dimensionless repulsion parameter of $a^* = 2\times10^6$. By simulating at lower numbers, one may extrapolate to the experimental values.

When the model is applied to granular solids, we may relate the particle-particle adhesion energy $G$ to the surface energy via $\gamma_s \sim G/d^2$. For dense granular solids, the volume fraction can be taken as a constant, the dense random packing, so that the number of bonds per particle is constant. Thus, up to a geometrical constant, $\gamma_s$ equals the energy that is needed to cleave the sample along a flat plane. Generally, a non-dented solid sample will fail under tension, compression or shear, at a maximum strain $\varepsilon_y$ that is proportional to the force range $\delta$. We checked this by simulation. From Eq. (6) we thus arrive at the prediction that the yield strain $\varepsilon_y$ of a granular material will follow

$$\varepsilon_y \propto \sqrt{\frac{\gamma_s}{Ed}} \quad (10)$$

Moreover, the yield stress $\sigma_y$ must scale as the maximum adhesion force divided by the particle area, hence $\sigma_y \propto \varepsilon/d^2$. Elimination of the force amplitude $\varepsilon$ using Eq. (9) leads to the prediction

$$\sigma_y \propto \sqrt{\frac{E\gamma_s}{d}} \quad (11)$$

These relations are quite remarkable results, as they coincide with the results derived from fracture mechanics [16]. Moreover this result is in line with a vast body of experimental work on fracture of granular solids loaded under uniaxial tension [17]. Eq. (11) is generally found for grain sizes smaller than a critical value, in which case the strength is controlled by crack propagation. Crack propagation in turn is related to viscous dissipation, which we describe below.

One may wonder why we do not use the Johnson-Kendall-Roberts (JKR) theory [18] for adhesion between colloidal particles. In the JKR theory the adhesion force is $F_{adh}^{JKR} = -3\pi\gamma_s R_i R_j/(R_i+R_j)$ where $\gamma_s$ is the surface energy of the particles. The consequence is that the yield stress is proportional to $\sigma_y^{JKR} \propto \gamma_s/d$. Hence, it depends on grain size and surface energy to a *wrong* power, and it is *independent* of the elastic modulus, whereas the dependence should be as in Eq. (11). Therefore the JKR theory is ruled out as a model for granular solids.

### B. Dissipative forces

Both for the simulation of colloidal particles, for powders and for granular solids, friction forces between neighboring particles are important. In general for granular matter, three modes of viscous damping can be distinguished: radial, tangential and torsional friction [19,20]. These modes are illustrated in FIG. 1. The first of these is the friction of particles on mutual approach, as the liquid in between is squeezed out (the squeeze mode). This is the friction mode that is included in standard DPD. This dissipative force is given by

$$\mathbf{F}_{ij}^D = -\gamma\xi(r_{ij})\mathbf{v}_{ij}^{//} = -\gamma\xi(r_{ij})(\mathbf{v}_{ij}\cdot\mathbf{e}_{ij})\mathbf{e}_{ij} \quad (12)$$

where $\mathbf{v}_{ij}$ is the velocity difference between the surfaces of neighboring particles $i$ and $j$; $\gamma$ is the friction factor associated to the radial (squeeze) mode; $\mathbf{e}_{ij}$ is a unit vector pointing from particle $j$ to particle $i$; and $\xi(r)$ is a distance dependent weight function, to be specified below.

The second mode of friction is a force perpendicular to the line of separation, and is caused by a shear force acting on surfaces that move tangentially (the shear mode). The last mode of friction is caused by rotation around the line of separation, which leads to a torque that damps out the rotational motion (the twist mode). A full account of the hydrodynamic forces between particles in close contact is given by Kim and Karilla [21]. For equally sized hard spheres, the squeeze and shear forces acting on a particle due to its neighbors are rather complicated. At small separation the leading contributions diverge with the separation distance. For pragmatic reasons we may however take a simplified expression that merely serves to introduce the right amount of viscous damping in the system.

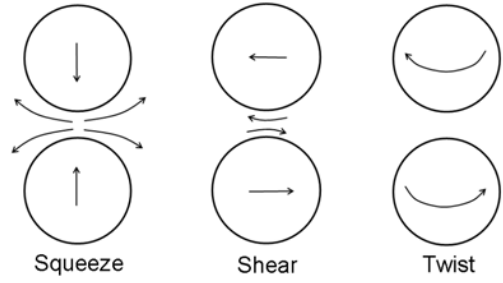

FIG. 1. Different modes of lubrication forces between particles.

Besides the squeeze mode given in Eq. (12), we will introduce the shear mode via a pair-wise force

$$\mathbf{F}_{ij}^S = -\mu\varsigma(r_{ij})\mathbf{v}_{ij}^\perp \quad (13)$$

where $\mathbf{v}_{ij}^\perp$ is the perpendicular velocity difference between the surfaces of neighboring particles; and $\mu$ is the shear friction factor. The function $\varsigma(r)$ is a distance dependent weight function. In practice, the friction forces are dominated by the squeeze mode; the shear mode is smaller, and we can ignore the twist mode.

The inclusion of non-central friction forces in a solid particle simulation was explored by Ball and Melrose [19] and for fluid particles by Español [22]. In this method all particles in the simulation have position $\mathbf{r}_i$, velocity $\mathbf{v}_i$ and spin $\boldsymbol{\omega}_i$. The equations of motion of the particles are given by

$$\begin{aligned}\dot{\mathbf{r}}_i &= \mathbf{v}_i \\ \dot{\mathbf{v}}_i &= \mathbf{F}_i/m \\ \dot{\boldsymbol{\omega}}_i &= \boldsymbol{\tau}_i/I\end{aligned} \quad (14)$$

where $\mathbf{F}_i = \Sigma_j \mathbf{F}_{ij}$ is the force acting on particle $i$ due to its neighbors. Similarly, the torque on particle $i$ is given by an interaction with the direct neighbors. If the torque is taken as

$$\boldsymbol{\tau}_i = -\tfrac{1}{2}\sum_j \mathbf{r}_{ij}\times\mathbf{F}_{ij} \quad (15)$$

where $\mathbf{r}_{ij} = \mathbf{r}_i - \mathbf{r}_j$, it follows immediately that the total angular momentum of the system is conserved [22].

Generally, the cause of friction between two shearing particles is the surface velocity difference. To obtain this, we calculate the tangent velocity of the particle surfaces at the point of closest contact. Thus for particles of arbitrary size we have

$$\mathbf{v}_i^\perp = \mathbf{v}_i - (\mathbf{v}_i \cdot \mathbf{e}_{ij})\mathbf{e}_{ij} - R_i \boldsymbol{\omega}_i \times \mathbf{e}_{ij}$$
$$\mathbf{v}_j^\perp = \mathbf{v}_j - (\mathbf{v}_j \cdot \mathbf{e}_{ij})\mathbf{e}_{ij} + R_j \boldsymbol{\omega}_j \times \mathbf{e}_{ij} \quad (16)$$

where particle radii are denoted by $R_i$ and $R_j$. Note that the sign in front of $\boldsymbol{\omega}$ is flipped for the $j$ particle because the vector from particle center to contact, points in the opposite direction as for particle $i$. The shear force is opposite to the tangential velocity difference; hence the shear force is in general given by

$$\mathbf{F}_{ij}^S = -\mu\varsigma(r_{ij})\mathbf{v}_{ij}^\perp = -\mu\varsigma(r_{ij})[\mathbf{v}_{ij} - (\mathbf{v}_{ij} \cdot \mathbf{e}_{ij})\mathbf{e}_{ij} - (R_i \boldsymbol{\omega}_i + R_j \boldsymbol{\omega}_j) \times \mathbf{e}_{ij}] \quad (17)$$

where $\mu$ is the shear friction factor.

### C. Random forces

To compensate for the energy loss by the radial friction and to simulate Brownian motion, a distance dependent random force is introduced. The general form for the radial random force is [2,23]

$$\mathbf{F}_{ij}^R = \sigma w(r_{ij}) \theta_{ij}(t) \mathbf{e}_{ij} / \sqrt{\delta t} \quad (18)$$

where $\theta_{ij}(t)$ is a random variable with zero mean and unit variance, and $\delta t$ is the time step used. The amplitude $\sigma$ and function $w(r)$ are related to the radial friction via the fluctuation-dissipation theorem [2,23]

$$\tfrac{1}{2}(\sigma w(r_{ij}))^2 = \gamma \xi(r_{ij})kT \quad (19)$$

This is the random noise as used in standard DPD. For the lateral modes, analogous relations have been derived by Ball and Melrose [19] and independently by Español [22], but these are quite complicated. A simpler alternative is investigated here.

We use a heuristic argument to find an appropriate noise function. The physical meaning of the fluctuation-dissipation theorem given in Eq. (19) is that at every distance from a particle center, an equal amount of kinetic energy is inserted in the form of random kicks, as is taken away by the friction force. The left hand side of the equation gives the inserted kinetic energy resulting from velocity change $\delta v_{ij} = \pm \sigma w(r_{ij})\sqrt{\delta t}$. The square of this is the energy input per unit of time, up to a factor of particle mass. The right hand side of Eq. (19) gives the amount of energy taken away per unit of time by friction, again up to a factor of particle mass, so that mass drops out of the equation. The amount of energy taken away by the shear friction is twice as much as for lateral friction, because there are two rotational degrees of freedom.

Thus we have the correct *energy balance* if we introduce radial noise with the local amplitude

$$\tfrac{1}{2}(\sigma w(r_{ij}))^2 = \gamma \xi(r_{ij})kT + 2\mu\varsigma(r_{ij})kT \quad (20)$$

More specifically, we can conveniently choose the radial and tangential weight functions to equal each other. The standard DPD choice is

$$\xi(r) = \zeta(r) = (w(r))^2 = (1 - r/r_c)^2 \theta(1 - r/r_c) \quad (21)$$

where $\theta(r)$ is the Heaviside step function, and $r_c$ is an appropriate cut-off distance. We will take $r_c$ = 1.5 unless stated otherwise. With this choice for the friction weight functions, the noise amplitude should be taken as

$$\sigma^2 = (2\gamma + 4\mu)kT \quad (22)$$

The method described here is not correct under all circumstances, because the energy dissipated in a lateral mode is reinserted in a radial mode. However, in test runs over $5 \times 10^4$ time steps for 1000 particles, for a sticky elastic sphere liquid at density $\rho d^3$ = 1, identical temperature, pressure and pair correlation were found as compared to DPD without lateral friction, see FIG. 2. Hence, equipartition is efficient enough to spread the inserted energy over all degrees of freedom, which ensures correct simulation. Thus it is concluded that the use of *only one* noise term is in practice sufficient to generate the correct thermodynamic ensemble, at least for central forces. This method is much simpler than the method introduced earlier [19, 22]. It is obvious that we regain standard DPD if we take $\delta = \varepsilon = \mu = 0$.

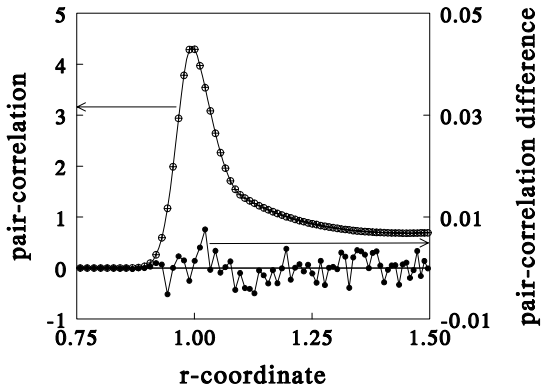

FIG. 2. Pair correlation for $a = 10^3$, $\delta = 0.1$ and $\varepsilon = 10$. The time step is $\delta t = 0.005$ and the density is $\rho d^3 = 1$. Crosses: $\gamma = \mu = 4.5$; circles: $\gamma = 4.5$, $\mu = 0$; dots: difference to enlarged scale. Noise is inserted radially, with amplitude given by Eq. (22).

When we introduce (conservative) bending forces between adjacent particles, however, an error is found in the temperature control if we also introduce shear friction ($\mu > 0$). In that particular case equipartition alone appears not to be efficient enough to spread the energy from the radial mode to the rotational degrees of freedom. Correct temperature control in the presence of bending forces is found when the shear viscosity is turned off ($\mu = 0$). The model including conservative shear and bending forces is clearly suitable to study elasticity problems, fracture mechanics and fragmentation of solids when noise is left out.

As an alternative to local friction and random noise, one may use another thermostat that preserves Galilean invariance. By construction this preserves inertial hydrodynamics [15,24] but it cannot be used at zero temperature. Therefore that method is not suitable for granular matter like powders, whereas the present method can be used at vanishing temperature.

To check the applicability to granular solids, the present simulation model has been used to simulate solid particle networks, where particles are bound together at their surfaces by (non-central) shear and bending forces, similar to the model proposed by Kun and Herrmann [25]. For colliding disks in 2D, Kun and Herrmann found a power law distribution of fragment sizes above a critical impact velocity [26].

For 3D systems the size distribution of fragments generally exhibits a power law with an exponential cut-off [27], in agreement with the experimental Gates-Gaudin-Schuhmann (GGS) distribution. Several authors have shown that the power laws are universal [28, 29, 30]. The present simulation model has been applied to the fragmentation problem in 3D, and we indeed found cluster size distributions in line with the GGS distribution.

In the remainder of this article we will concentrate on the phase behavior of the model without conservative bending and shear forces. This is obviously relevant for colloids but also for powders, because of the analogy between powders and liquids made by Edwards [6]. This analogy led to the notion that in dense granular media a vibration temperature – or granular temperature – can be always defined [12,31] although the interpretation may not be obvious. In particular, very different regimes of granular media have been distinguished that correspond to glassy, liquid and granular gas phases [12]. To map out at what granular temperature these regimes occur, and how this depends on the interaction range, we have studied the phase diagram of the present Sticky Elastic Sphere model.

### III. PHASE DIAGRAM

To analyze the Sticky Elastic Sphere phase diagram we first recall the Adhesive Hard Sphere (AHS) model that was introduced by Baxter [32] in 1968. This consists of hard spheres with an attractive interaction of vanishing range. The interaction potential is

$$\exp(-U(r)/kT) = \theta(r-d) + \frac{d}{12\tau}\delta(r-d)$$

(23)

where $\theta(r)$ is the Heaviside step function that accounts for the hard core repulsion, and the Dirac $\delta(r)$ distribution represents adhesion. The parameter $\tau$ takes on the role of temperature, but it is actually related to an association constant. Using the Percus-Yevick closure, Baxter came up with an analytical solution to the equation of state which indicates liquid-vapor coexistence below a particular critical temperature.

Because the Percus-Yevick theory is not exact, two estimates for the critical point widely differ. Simulating the model in 3D is quite challenging because interaction only takes place at contact. It took until 2003 that it was directly modeled with Monte Carlo simulation. Miller and Frenkel [14,33] obtained the critical point as $\tau^c$ = 0.1133±0.0005 and $\rho^c d^3$ = 0.508±0.01.

If we use this model as a reference, we can estimate the properties of the Sticky Elastic Sphere model as function of the adhesion energy and as function of the range of the attractive potential. To this end we first calculate the association constant between two spheres. This is given by the excess of the 2$^{nd}$ virial coefficient over the purely repulsive interaction [34,35,36] (see Eq. (4)), and follows as

$$K_a = 4\pi \int_0^{(1+\delta)d} [\exp(-U(r)/kT) - \exp(-U^{Rep}(r)/kT)] r^2 dr \approx 4\pi d^3 \delta \sqrt{\frac{\pi kT}{3G}} \exp(G/kT) \quad (24)$$

From the definitions of $\tau$ and $K_a$ we identify $K_a = 4\pi d^3/12\tau$. If the range $\delta$ is sufficiently small thermodynamics is determined by the association constant only. In that case the Sticky Elastic Sphere model should follow the AHS model. This means that the critical point follows from $K_a^c = 4\pi d^3/12\tau^c = 9.24\pm0.04$. Expressing the association constant in terms of adsorption energy and numerically solving $K_a = K_a^c$, we find the critical adsorption energy as function of the interaction range. For a force range between $10^{-4} < \delta < 0.3$ this critical association energy is very well described by the following function:

$$G/kT_{AHS}^c \approx -\ln(\delta) + 1.0 - 1.5\delta^{0.2} + 1.1\delta \quad (25)$$

The leading term $-\ln(\delta)$ follows from the analytical approximation in Eq. (24). The subscript AHS indicates that this result is obtained by mapping the Sticky Elastic Sphere model onto the AHS model. Note that the association constant was calculated by numerical integration, because the analytical approximation given in Eq. (24) has small but systematic errors. Based on the AHS results, we might thus expect liquid-vapor coexistence for sticky elastic spheres when the adhesion energy is larger than this critical value.

However, the validity of this result may be questioned. Miller and Frenkel [14,33] indeed find liquid-vapor coexistence for an interaction potential of vanishing range. But on the other hand, Hagen and Frenkel [13] found that if the interaction range in a Yukawa fluid is shorter than about 1/6$^{th}$ of the particle diameter, liquid-vapor coexistence disappears, and only solid-vapor coexistence is found. Further analytical theory [37] and simulations [38] indicate that the liquid phase cannot be stable for a force range smaller than 1/6.

To investigate this apparent contradiction, simulations were done for systems comprising 2000 particles in a box of size 12.5×12.5×25. The overall mean density is $\rho d^3$ = 0.512, which is within the error bar of the estimated critical density $\rho d^3$ = 0.508±0.01, as reported by Miller and Frenkel [14,33]. Temperature, repulsion parameter, and force range were varied using $a = 6/\delta^2$ and $\varepsilon = 1.5/\delta$. This describes elastic spheres with a fixed adhesion energy $G = 1$. Time steps used are specified in Table I. In simulations described below we used a radial friction factor $\gamma = 10$, shear friction $\mu = 0$; and to simulate Brownian motion noise was added according to Eq. (22). In all cases the error in the temperature control was less than 0.5%.

We start with a conformation where all particles are placed in the left half of the box. For a force range $\delta = 0.2$ the system quickly establishes a liquid-vapor coexistence for $kT = 0.5$, 0.53 and 0.55, whereas large fluctuations are seen at $kT = 0.6$. For this force range the mapping to adhesive hard spheres predicts a critical point at $kT^c_{AHS} = 0.576$, which is consistent with the observations. For the smaller range $\delta = 0.1$, the mapping to the AHS model predicts $kT^c_{AHS} = 0.406$, and indeed we find large fluctuations at $kT = 0.41$. This suggests that a critical point is nearby. However, for the slightly lower temperature $kT = 0.4$, the system immediately crystallizes and we find solid-vapor coexistence, see FIG. 3. This indicates that the liquid phase is indeed unstable for $\delta = 0.1$. The question thus arises, when do we have liquid-vapor coexistence, when solid-vapor and when solid-liquid?

If a liquid or vapor phase has a lower free energy than the solid phase, a solid will immediately melt. Reversely, it is difficult to nucleate a solid from a meta-stable liquid or vapor. Therefore we started with a solid-vapor coexisting system, and changed the force range and temperature to explore new points in the phase diagram. Generally, for low temperatures we find a stable crystalline phase. Increasing the temperature for small force ranges ($\delta < 0.19$), the crystal is observed to dissolve into the vapor phase *without* first melting into a liquid. In contrast, for a larger force range ($\delta > 0.19$) we find the solid phase to melt if $T_m < T < T^c$, and liquid-vapor coexistence disappears above critical temperature $T^c$, see FIG. 4.

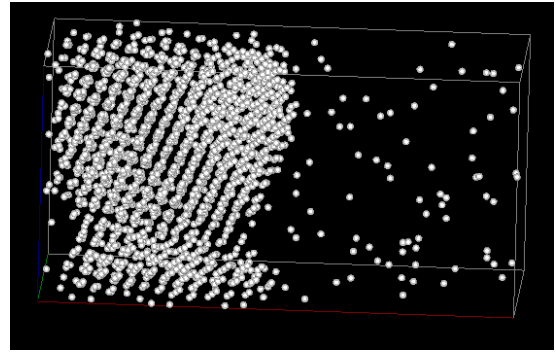

FIG. 3. A typical configuration obtained for $kT/G = 0.4$, and $\delta = 0.1$. The time step is $\delta t = 0.01$ and we used friction factors $\gamma = 10$ and $\mu = 0$. Noise is inserted radially, with amplitude given by Eq. (22). The left part of the system is a poly-crystalline solid; the right-hand side is a vapor phase. For clarity the particles are drawn with radius $R = 0.2$, in reality the simulated radius is $R = 0.5$.

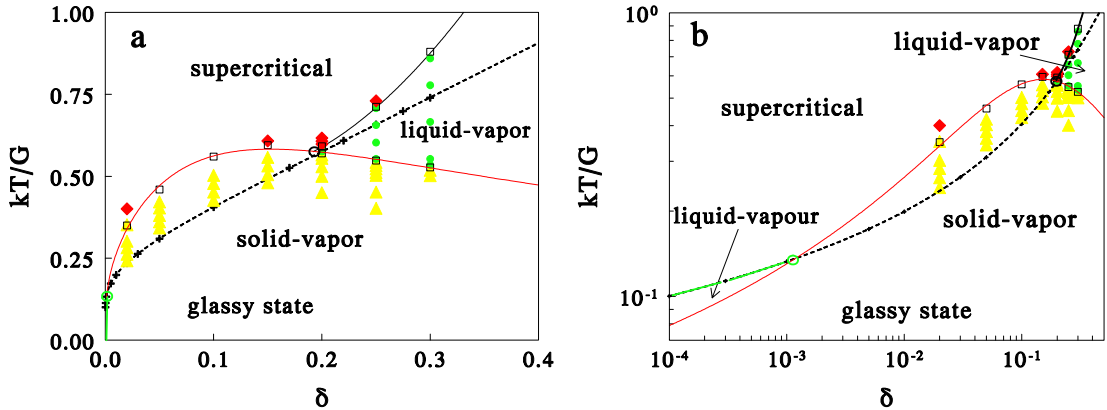

FIG. 4. (Color online) (a) Approximate phase diagram of sticky elastic spheres near critical density. The dashed curve with black crosses is the line of liquid-vapor critical points, as calculated from the AHS model. The full curves are melting line (red) and line of critical points (black) estimated from the simulations. Yellow triangles denote solid-vapor coexistence, green dots indicate liquid-vapor coexistence and red diamonds are super critical. The black squares are the estimated melting and critical points. The green and black circles denote critical end-points where the lines of critical points and the melting transition line meet. The dashed line between these points is a line of hidden critical points where crystallization will be relatively fast. A glass transition line is present below the melting line, but is not drawn in. (b) Same diagram in log-log scale, showing extrapolation to small force range.

To find the critical point for each force range, the surface tension $\sigma$ was measured at various temperatures. Close to the critical point this is expected to decrease as $\sigma \propto (1-T/T^c)^\mu$ where the exponent $\mu = 1.26$ is pertinent for Ising criticality in 3D [2]. Mean-field theory predicts $\mu = 1.5$. For short-range interaction potentials, the exponent should be universal, only for very long range interactions one may expect mean-field criticality, as Ising criticality is then limited to a narrow temperature range. It was indeed observed that the Ising exponent $\mu = 1.26$ fits the present simulation results quite well, see FIG. 5. Finite size effects may play a role very close to the critical point, but extrapolation of the surface tension to zero still gives a reasonable estimate of the critical temperature, as for most systems the width of the interface is small compared to the system size. For $\delta = 0.2$, however, because of the narrow temperature range between the melting point and critical point, eleven systems were simulated in the temperature range from $kT = 0.45$ to $kT = 0.62$, and the melting point and critical point were estimated from the observed phase behavior. The critical points are plotted in the phase diagram, FIG. 4. The line of liquid-vapor critical points is found to intercept the melting line roughly at $\delta = 0.19$.

This implies that below $\delta = 0.19$ there is no thermodynamically stable liquid phase. For this force range the width of the peak in the Mayer function at half height is 0.154. This compares well with the interaction range of the Yukawa fluid (roughly 1/6) where the liquid phase disappears [13,37].

TABLE I. Simulation parameters and estimated melting point and critical temperature at mean density $\rho = 0.512$. The last column gives the critical temperature as calculated from the mapping to the Adhesive Hard Sphere model. Attraction and repulsion parameters are chosen as $\varepsilon = 1.5/\delta$ and $a = 6/\delta^2$, corresponding to adhesion energy $G = 1$.

| $\delta$ | $\varepsilon$ | $a$ | $\delta t$ | $kT_m/G$ | $kT^c/G$ | $kT^c_{AHS}/G$ |
|---|---|---|---|---|---|---|
| 0.02 | 75 | 15000 | 0.001 | 0.35±0.01 | - | 0.2351 |
| 0.05 | 30 | 2400 | 0.005 | 0.46±0.01 | - | 0.3097 |
| 0.10 | 15 | 600 | 0.01 | 0.56±0.01 | - | 0.4059 |
| 0.15 | 10 | 266.67 | 0.02 | 0.60±0.01 | - | 0.4925 |
| 0.20 | 7.5 | 150 | 0.02 | 0.570±0.005 | 0.592±0.006 | 0.5758 |
| 0.25 | 6 | 96 | 0.02 | 0.548±0.005 | 0.713±0.005 | 0.6580 |
| 0.30 | 5 | 66.67 | 0.02 | 0.527±0.005 | 0.88±0.01 | 0.7398 |

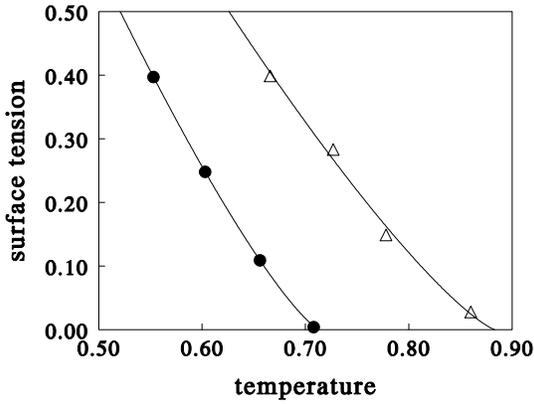

FIG. 5. Surface tension as function of temperature for force ranges $\delta = 0.25$ (dots) and $\delta = 0.3$ (triangles). Lines are fits to Ising criticality, $\sigma \propto (1-T/T_c)^{1.26}$.

As we go down in force range it becomes more difficult to establish the melting point, as dynamics are slow and moreover we need to take small time steps. Therefore, runs of 120,000 steps were used in duplicate to check equilibration. To establish the melting temperature, we measured the vapor density in coexistence with the solid phase. The vapor density increases exponentially with temperature, either as $\ln(\rho_v) \propto a+bT$ or as $\ln(\rho_v) \propto a-b/T$. In the limit of infinite system size the crystal must melt at the temperature where the (extrapolated) density of the vapor phase equals the mean density of the system ($\rho = 0.512$). Thus, the vapor density as function of temperature was extrapolated to $\rho = 0.512$; the corresponding temperature is our estimate of the melting temperature at this density.

Close to $\delta = 0$ the partition function of the solid phase can be estimated roughly as $Z_N \sim (\delta^3 \exp(G/kT))^N/N!$, hence the solid state chemical potential should diverge with the force range as $\mu \sim -G/kT - 3\ln(\delta)$. Adding an empirical linear term and a similar power law of $\delta$ as in Eq. (25), we find that the melting line is fitted well by

$$G/kT_m \approx -3\ln(\delta) - 18.2 + 20.5\delta^{0.2} + 1.2\delta \quad (26)$$

The melting curve is shown in FIG. 4, together with the critical line that is based on mapping to the AHS model. Simulation parameters and estimated melting temperature at density $\rho = 0.512$, and the liquid-vapor critical points are summarized in Table I.

Because the melting point $G/kT_m \sim -3\ln(\delta)$ diverges slightly faster than the liquid-vapor critical point, $G/kT^c_{AHS} \sim -\ln(\delta)$, for very small force range the melting transition must cross the critical line. Hence, for very small force range the solid phase becomes unstable for entropic reasons – the spheres have no room to move in a solid – and the system has liquid-vapor coexistence again below the critical temperature. Based on the best estimate of the melting line Eq. (26) and the mapping to the AHS model Eq. (25), the simulations predict a re-entrant liquid phase for a force range below $\delta \approx 10^{-3}$ and an adsorption energy $G/kT \approx 7.5$. This indicates that the adhesive hard sphere model ($\delta \to 0$) does have a stable liquid phase, but this is a rather pathological limit; a stable solid phase is missing. The re-entrant liquid phase is indicated in the phase diagram, see FIG. 4b. Because the AHS model has no solid phase, the Sticky Elastic Sphere model is for most applications more realistic.

The lines of critical points, which demarcate the super critical and liquid-vapor regions, meet the (first-order) melting transition line. It is generally acknowledged that upon meeting a first-order phase transition line, a line of critical points must end in either a tricritical point or a critical end-point. The latter case is relevant here. Such critical end-points are indicated in FIGs 4a and 4b by the green and black circles. The dashed line between the two critical end-points presents a line of hidden critical points at each value of the force range, where relatively fast crystallization is expected [39]. This line is based on a perturbation from the Adhesive Hard Sphere model. The large fluctuations and rapid crystallization observed at parameters $\delta = 0.1$ and $kT/G = 0.4$ are consistent with this prediction.

When a system is quenched to $T = 0$ (leaving out the random noise), friction gradually slows down all particle motion. In general it is observed that $T = 0$ can be reached before a crystalline solid is formed. For example, a system at density $\rho d^3 = 1.16$ was first equilibrated at $kT = 2$ and then quenched to $T = 0$ using $\gamma = 10$, $\mu = 0$, and $r_c = 3$. FIG. 6 shows the pair-correlation functions obtained for simulation parameters $\delta = 0.1$, $\varepsilon = 15$ and $a = 600$ at $kT = 2$, and for the quenched system at $T = 0$. The high temperature system clearly has a liquid pair-correlation, whereas the quenched system has extra peaks. The triangular symbols are markers at distance 1, $\sqrt{2}$, $\sqrt{3}$, etc. Thus, the quenched system shows a peak at $r = \sqrt{3}$, but all other structure characteristic of a crystalline phase is missing. Since all evolution has come to a standstill, it follows that the quenched system must be a glass. Indeed, the pressure and temperature relax by stretched exponentials, P ~ $\exp(-(t/t_0)^{0.3})$, which is characteristic of a glassy state. This shows the existence of a glass transition line in the phase diagram below the melting curve, but its precise location as function of temperature, force range and friction factor has not been determined.

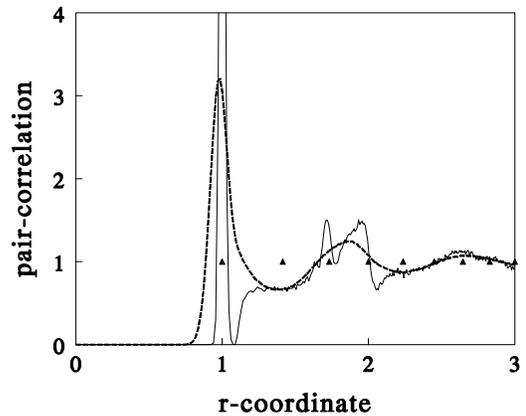

FIG. 6. Pair-correlation for a system at $kT = 2$ (dashed curve), and a quenched system at $T = 0$ (full curve). The triangles are markers at distance 1, $\sqrt{2}$, etc. Simulation parameters are given in main text.

## IV. DISCUSSION AND CONCLUSIONS

A very simple simulation model for adhesive solid particles is described. This model is well suited to describe the interaction between colloidal particles, granular powders and granular solids. The interaction force increases linearly within a contact distance, and outside this distance it is represented by a parabola over a small but finite range. The interaction range and amplitude are related to the elastic modulus of the particles and to the particle-particle adhesion energy or fracture energy. When applied to granular solids, the correct scaling rule for the fracture stress as function of surface energy, elastic modulus and grain size is obtained. In contrast, the Johnson-Kendall-Roberts theory for adhesion between colloidal particles, predicts the yield stress to be independent of the particle modulus, and to depend on particle size and surface energy to wrong powers.

To incorporate viscous interactions, radial and shear frictions are added between neighboring particles, and radial noise is added to simulate Brownian motion at a finite temperature. Radial noise leads to correct temperature stabilization and to the correct pair-correlation function – even if shear friction is added – when the noise amplitude is increased to compensate for the viscous drag of rotational degrees of freedom. Only when particles interact via (non-central) bending forces, temperature control by radial noise fails when shear friction is included.

To investigate the model at finite (granular) temperatures, the phase diagram as function of temperature and interaction range has been determined for a fixed (near critical) density. Liquid-vapor coexistence is found for sufficiently large interaction range at an intermediate temperature range. At lower temperature the solid phase is stable, and at higher temperature liquid-vapor coexistence ends at the critical temperature. At temperatures below the melting line, the simulation model forms a glass, with characteristic stretched exponential pressure decay. For varying force range, the line of critical temperatures hits the melting line at an interaction range of 19% of the particle diameter at a critical end-point. For shorter interaction ranges the liquid phase ceases to exist, in line with known results for the Yukawa fluid.

For very small interaction range the free energy of the solid phase and of the liquid phase both diverge proportional to $\ln(\delta)$. However, because the free energy of the solid diverges faster than that of the liquid, the critical line must cross the melting line again at a critical end-point at very short interaction range, which is estimated as $\delta \approx 10^{-3}$. For shorter force ranges a stable liquid phase reappears. The use of Baxter's Adhesive Hard Sphere model to describe the structure of colloidal dispersions or wet powders is thus strictly allowed only if the interaction range is (well) below 0.1% of the particle diameter.